# The Dark Matter Distribution in Galaxy Cluster Cores

J.S. Arabadjis[*], M.W. Bautz[†] and G. Arabadjis[**]

[*]*Massachusetts Institute of Technology, Center for Space Research,
70 Vassar Street, Room 37-501, Cambridge, MA 02139*
[†]*Massachusetts Institute of Technology, Center for Space Research,
70 Vassar Street, Room 37-521, Cambridge, MA 02139*
[**]*Mitre Corporation, 202 Burlington Road, Bedford, MA 01730*

**Abstract.**
Determining the structure of galaxy clusters is essential for an understanding of large scale structure in the universe, and may hold important clues to the identity and nature of dark matter particles. Moreover, the core dark matter distribution may offer insight into the structure formation process. Unfortunately, cluster cores also tend to be the site of complicated astrophysics. X-ray imaging spectroscopy of relaxed clusters, a standard technique for mapping their dark matter distributions, is often complicated by the presence of their putative "cooling flow" gas, and the dark matter profile one derives for a cluster is sensitive to assumptions made about the distribution of this gas. Here we present a statistical analysis of these assumptions and their effect on our understanding of dark matter in galaxy clusters.

## Introduction

The cold dark matter (CDM) paradigm of modern cosmology has enjoyed spectacular success in describing the formation of large-scale structure in the universe [1, 2, 3, 4]. Galaxy-scale dark matter halos, however, exhibit several apparent inconsistencies with CDM, for example: the number of Milky Way satellites appears to be at least an order of magnitude lower than CDM predictions [5, 6, 7], and dark matter halos in dwarf and low surface brightness galaxies are much less cuspy than in CDM simulations [8, 9, 4]. Some reports [11, 12] even suggest that CDM fails on galaxy cluster scales for some clusters, but the latter are controversial [13, 14].

If CDM does indeed require alterations, there is no shortage of tailors. Proposed modifications include, though are not limited to, self-interacting dark matter [15, 16], warm dark matter [17], annihilating dark matter [18], scalar field dark matter [19, 20], and mirror matter [21], each of which is invoked to soften the core density profile. Many of these modifications will soften the core profile of galaxy clusters as well, although other astrophysical processes such as the adiabatic contraction of core baryons [22] may ameliorate this effect. Baryons, however, introduce a host of complications to CDM simulations; their effects will require a great deal of effort to disentangle [23].

In order to discriminate among CDM, its modifications, and other astrophysical influences, we have initiated a program to map the dark matter profiles of a sample of galaxy cluster cores. We use imaging spectroscopy from *Chandra X-ray Observatory* obser-vations [24] and archival data [25] to extract the deprojected radial dependence of the baryon density and temperature of each cluster. In order to extract a dark matter profile from spatially resolved X-ray spectroscopy one usually assumes that the galaxy cluster is spherically symmetric and in hydrostatic equilibrium, and so for the most part we have restricted our sample to clusters for which these assumptions are most likely valid. Unfortunately, these clusters often contain "cooling flow" gas in their cores which complicates the spatial and spectral models. How one models the X-ray emission from this cool gas can significantly alter the resulting dark matter profile. If the model contains only a single emitting component (at temperature $T$ and density $\rho$) at each radius, the inferred temperature profile will tend to dip significantly toward the center of a cooling flow cluster. If, however, gas in the the central few radial bins contains a second (cooler) component which is cospatial and isobaric with the first, the hot gas temperature profile remains flat into the core. The latter case tends to produce a larger central mass (see Figure 1A) and steeper density profile than the the former. Our problem, then, is the age-old exercise of choosing between a simple model $\mathsf{M}^\mathsf{s}$ and a complex model $\mathsf{M}^\mathsf{c}$. Once we have done this, we will be better able to address the other issues listed above.

## Models

In order to ascertain the importance of a second emission component we adopt a simplified geometry containing only two spherical shells (inner = 1, outer = 2). In both models ($\mathsf{M}^\mathsf{s}$ and $\mathsf{M}^\mathsf{c}$), shell 2 contains a (hot) thermal plasma at temperature $T_{2h}$ and density $\rho_{2h}$. Model $\mathsf{M}^\mathsf{s}$ contains only one emission component in shell 1, characterized by a temperature $T_{1h}$ and a density $\rho_{1h}$, whereas model $\mathsf{M}^\mathsf{c}$ contains a hot and a cool emission component in shell 1, described by $T_{1h}$, $\rho_{1h}$, $T_{1c}$, and $\rho_{1c}$. The X-ray emission from each component is modelled spectroscopically as using the MEKAL [26, 27, 28, 29] model as implemented in the XSPEC software package [30]. The best-fit parameter values of each model are calculated using a $\chi^2$ minimization routine. Hereafter we will refer to the simple and complex model parameters using vectors $\theta^\mathsf{s}$ and $\theta^\mathsf{c}$, respectively; i.e., $\theta^\mathsf{s} = (T_{1h}, \rho_{1h}, T_{2h}, \rho_{2h})$ and $\theta^\mathsf{c} = (T_{1h}, \rho_{1h}, T_{1c}, \rho_{1c}, T_{2h}, \rho_{2h})$.

At this point, tradition dictates that we employ a statistical test from the standard arsenals [31, 32, 33], such as the likelihood ratio test or the $F$-test, to choose between the two models. However, since $\theta^\mathsf{s}$ lies on a boundary of $\theta^\mathsf{c}$ (with $\rho_{1c} = 0$), these tests cannot be employed [34]. Instead, we construct an *empirical F-distribution* using Markov Chain Monte Carlo (MCMC) sampling, and gauge the significance of the complex model from the location of the $F$ value of the data within that distribution [34].

## Constructing an empirical $F$-distribution

Starting with the best-fit parameters $\theta_0$ of model $\mathsf{M}^\mathsf{s}$, we sample the 4D parameter space in its vicinity using MCMC sampling. This is done by running a Tcl script within XSPEC which calculates the probability distribution function $P$ of a trial perturbation $\theta_1$ about $\theta_0$ given the observed data. The trial point is chosen using the trial distribu-tion function $q(\theta_0, \theta_1)$. The choice of $q$ is arbitrary; we restrict ourselves to functions which are symmetric in parameter space transitions, i.e. $q(\theta_i, \theta_j) = q(\theta_j, \theta_i)$ (this is the Metropolis algorithm – see `ftp://ftp.cs.utoronto.ca/pub/radford/review.ps.Z`; in this example we used a 4D gaussian deviate). This new parameter set is accepted if $P(\theta_1)/P(\theta_0)$ exceeds a random number on [0,1]. If not, the trial point is rejected and new one is selected. The sequence of accepted $\theta_i$ is a Markov Chain whose stationary distribution follows $P(\theta)$ [35]. We repeat this procedure until we have 100 values of $\theta$ for model $M^s$.

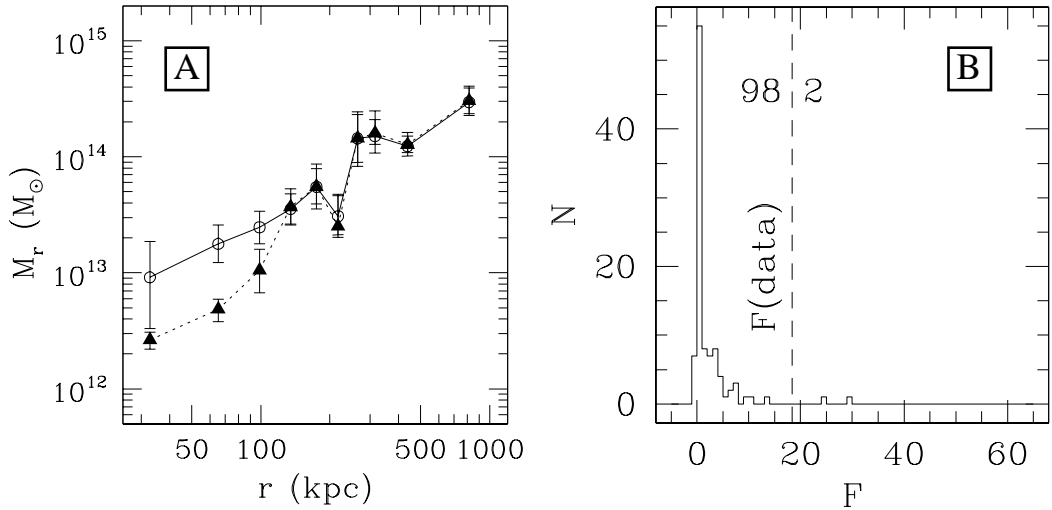

**FIGURE 1.** [A] One- and two-temperature mass profiles (dotted/triangles and solid/circles, respectively) of EMSS 1358+6245; [B] empirical $F$-distribution for models $M^s$ and $M^c$ of EMSS 1358+6245.

For each of the parameter sets in the sample we simulate an X-ray spectrum, taking proper account of the instrument response and photon statistics. We then model each of the simulated data sets using both $M^s$ and $M^c$, and tabulate the $F$-value of each data set:

$$F = \frac{\chi^2(\theta^s) - \chi^2(\theta^c)}{\chi^2(\theta^s)/\nu(M^s)} \quad (1)$$

where $\nu(M^s)$ is the number of degrees of freedom of the simple model. (In practice, the normalization can be ignored.) The $F$-distribution for the cooling flow cluster EMSS 1358+6245 is shown in Figure 1B. The $F$-value of the original *Chandra* data set is indicated with a dashed line.

## Conclusion

Of the 100 MCMC simulations that were run, only two resulted in an $F$-value which exceeded that of the data – that is, if $M^s$ were the correct description, an $F$-value as large as that observed would occur with a probability of only 2% – meaning that the model with a separate, co-spatial cool component is preferred. The result is that a model with a steeper density profile and a larger central mass is favored. If this trend obtains for most cooling flow clusters, it may rule out several of the CDM modifications.

We have demonstrated a technique which provides a rigorous and quantitative procedure for deciding between emission models of cooling flow clusters. As we continue to model clusters in the *Chandra* archive, this method will faciliate comparisons with CDM simulations by helping to remove much of the uncertainty in the derivation of cluster core density profiles.